# The habitat of early life:

# Solar X-ray and UV radiation at Earth's surface 4-3.5 billion years ago.

Ingrid Cnossen[1,2,] Jorge Sanz-Forcada[1,3], Fabio Favata[1,] Olivier Witasse[1], Tanja Zegers[1,4,] Neil F. Arnold[2]

[1]Research and Scientific Support Department of ESA, PO Box 299, 2200 AG Noordwijk, The Netherlands
[2]Radio and Space Plasma Physics Group, University of Leicester, University Road, Leicester, LE1 7RH, United Kingdom
[3]Now at: Laboratory for Space Astrophysics and Theoretical Physics (LAEFF), PO Box 50727, 28080 Madrid, Spain
[4]Now at: Faculty of Geosciences, Utrecht University, PO Box 80115, 3508 TC Utrecht, The Netherlands

**Abstract**

Solar X-ray and UV radiation (0.1-320 nm) received at Earth's surface is an important aspect of the circumstances under which life formed on Earth. The quantity that is received depends on two main variables: the emission of radiation by the young Sun and its extinction through absorption and scattering by the Earth's early atmosphere. The spectrum emitted by the Sun when life formed, between 4 and 3.5 Ga, was modeled here, including the effects of flares and activity cycles, using a solar-like star that has the same age now as the Sun had 4-3.5 Ga. Atmospheric extinction was calculated using the Beer-Lambert law, assuming several density profiles for the atmosphere of the Archean Earth. We found that almost all radiation with a wavelength shorter than 200 nm is attenuated effectively, even by very tenuous atmospheres. Longer wavelength radiation is progressively less well attenuated, and its extinction is more sensitive to atmospheric composition. Minor atmospheric components, such as methane, ozone, water vapor, etc., have only negligible effects, but changes in $CO_2$ concentration can cause large differences in surface flux. Differences due to variability in solar emission are small compared to this. In all cases surface radiation levels on the Archean Earth were several orders of magnitude higher in the 200-300 nm wavelength range than current levels in this range. That means that any form of life that might have been present at Earth's surface 4-3.5 Ga, must have been exposed to much higher quantities of damaging radiation than at present.

# 1. Introduction

Our understanding of the origin of life, as well as its early evolution, is still quite limited, despite decades, if not centuries of active research (see e.g. the review by Oró et al., 1990). In order to unravel part of the mystery, many studies have focused on the circumstances under which life formed, since we would gain a better insight in the nature of early life itself, if we knew more about its habitat. The aspect of the habitat of early life that is addressed here is the X-ray and UV radiation that was received at Earth's surface 4-3.5 Ga. Researchers have been interested in this topic for almost half a century, starting with Sagan [1957; 1961], and followed by many others [e.g. Berkner and Marshall, 1965; Kasting et al., 1989; Cockell, 1998, and references therein], since this aspect of the environment is very relevant to early life.

First of all, it has been proposed that X-ray and/or UV radiation could have provided the energy required for the chemical reactions that resulted in the formation of (more complex forms of) organic material, eventually leading to the origin of life [e.g. Miller and Urey, 1959]. However, on living tissue, the same radiation has a damaging effect. It can, for example, create mutations in DNA molecules that can be fatal to the organism. On the other hand, these mutations do create the genetic variation which is required for evolution to take place. In any case, whether positive or negative, X-ray and UV radiation must have played a key role in the conditions under which life formed and evolved.

The exact timing of the formation of life on Earth is uncertain, but it is usually placed between 3.8 and 2.7 Ga. The evidence for an early origin of life, around 3.8 Ga, consists of the presence of isotopically light carbon, suggestive of biological processes, in banded iron formations belonging to the ~3.8 Ga Isua and Akilia belts in Greenland, [Mojzsis et al. 1996; Rosing, 1999; Schidlowski, 2001]. However, recent studies [Fedo and Whitehouse, 2002; Zuilen et al., 2003; Lepland et al., 2005] indicate that the minerals that contain the isotopically light carbon may be much younger than 3.8 Ga. Evidence for the presence of life 3.5 Ga is more abundant and varied, consisting of (micro)fossils of primitive single-cell organisms, stromatolites, as well as isotopic evidence [e.g. Altermann and Kazmierczak, 2003 and references therein]. The present study focuses on the conditions at Earth's surface in the time window from 4 to 3.5 Ga, relevant for an early origin of life.

4 to 3.5 billion years ago, the Sun was likely to be less bright than today: calculations based on a standard solar model estimate its bolometric luminosity in the Archean at 74-77% of its present value [Bahcall et al., 2001]. In contrast, the solar X-ray and UV luminosity emitted by the outer atmosphere was likely much higher in the past, by analogy with other (solar-like) stars that generally show a decrease in X-ray and UV luminosity with age [Micela et al., 1985, 1988; Micela, 2002]. This is related to the relatively fast rotation of young stars, which in the case of Sun-like stars, yields to a magnetic dynamo generated activity. This activity is responsible for a high emission in the X-ray, EUV and FUV bands, which decreases with the age of the star, as its rotation progressively slows down. Therefore, much higher fluxes of short-wavelength radiation were released at the top of Earth's atmosphere in the Archean than at present. The amount of radiation that actually reached the surface depended on the composition of the atmosphere 4-3.5 Ga.

Earth's early atmosphere was different from the current atmosphere [see e.g. Kasting, 1993; Frimmel, 2005, and references therein]. The most dominant constituent of Earth's atmosphere, nitrogen, was probably present in similar (mass) concentrations as it is today [e.g. Selsis, 2004], since there is only little exchange of this gas between the atmosphere and the solid Earth, and because nitrogen hardly reacts with other components in the atmosphere. However, Earth's atmosphere must have changed considerably in other aspects: carbon dioxide, oxygen, and ozone concentrations for instance were probably very different from today.

Higher concentrations of greenhouse gases must have been present in Earth's early atmosphere, to solve the so-called "faint young Sun paradox" [Sagan and Mullen, 1972]. This refers to the apparently paradoxal evidence for the presence of liquid water (as clearly indicated by geological evidence such as pillow basalts), while the Sun did not provide sufficient energy for water to be present at the Earth's surface in any other form than ice, under present-day atmospheric conditions. Although other possible solutions have been proposed, e.g. a more massive young Sun which has lost mass through a significant wind, thus having a mass decreasing with time but a constant luminosity [Sackmann & Boothroyd, 2003], it is generally thought that the atmospheric composition of the early Earth was different, including more greenhouse gases to warm the Earth's surface and allow the presence of liquid water. Proposed greenhouse gases for Earth's early atmosphere include carbon dioxide, methane, ammonia, and water vapor. However, it is difficult to justify high concentrations of methane and ammonia in a pre-life stage, since non-biological sources are not sufficiently available, and especially ammonia quickly photodissociates under the action of UV radiation [Kuhn and Atreya, 1979].

The Earth's early atmosphere probably contained only very low concentrations of oxygen, if at all [Kasting, 1993; Frimmel, 2005], whereas it constitutes more than 20% of the current atmosphere. Only from 2.2-2 Ga onward oxygen levels started to increase [Rye and Holland, 1998] as a by-product of biological activity. Inorganic sources for $O_2$ include photolysis of $CO_2$ and/or $H_2O$, which might be considered important in an atmosphere containing high concentrations of $CO_2$, but still these were not sufficient to build up oxygen levels significantly in the Archean [Wayne, 2000, p.666]. Due to the low oxygen content, concentrations of ozone were likely much lower as well, as ozone is formed by photochemical reactions from oxygen. Ozone is known to be an important shielding component for UV radiation in the present atmosphere.

Previous studies on the UV radiation reaching Earth's surface in the Archean have been carried out relatively recently by Cockell and collaborators [Cockell, 1998, 2000a/b; Cockell et al., 2000]. They covered a wavelength interval of ~200-400 nm, and found that in this range radiation levels at the Archean surface were much higher than today. They assumed that the Sun had 75% of its present brightness over the complete wavelength range and that the atmosphere did not contain any oxygen or ozone.

In the present study, we modeled the spectral radiance of the young Sun based on the evolution of solar-like stars. This allowed us to model the young Sun's spectrum for shorter wavelengths (from 0.1 nm onward). In addition, we assessed effects of the uncertainty in the activity level of the early Sun and effects of activity cycles and flares. Also, to determine the solar flux at the surface of the early Earth, we analyzed the sensitivity of the surface spectral irradiance with respect to atmospheric composition. This included tests with different levels of carbon dioxide, and small amounts of methane, water vapor, and ozone. Previous studies generally used only one typical atmosphere, and therefore ignored any dependence of surface fluxes on atmospheric composition.

We will show that, despite considerable changes in the solar emission with respect to present-day, the solar flux at the surface of the early Earth was largely determined by the composition of the atmosphere, while changes in solar emission had only minor effects. Comparison between the present and Archean surface irradiance will show that Archean radiation levels were much higher, in a wavelength range to which life is particularly sensitive.

**Methods**

*2.1 Solar spectral irradiance 4-3.5 Ga*

As the Sun has evolved with time, the young Sun had a different effective temperature than the present-day Sun, which resulted in a different emitted spectral energy distribution (SED). The present-day Sun is a G2 star and has an effective temperature of 5777 K. In order to model the SED of the young Sun, we have compared it to the G5 star $\kappa^1$ Cet (HD 20630), with an effective temperature of 5600 K. $\kappa^1$ Cet is a solar-like star, meaning that it has a similar mass, size and metallicity as the Sun, but it has an age of approximately 0.5 Gyr [Lachaume et al., 1999]: the same age as the Sun had 3.95-3.85 Ga. Therefore $\kappa^1$ Cet is an appropriate proxy for the young Sun for the wavelength range of interest here.

The UV and X-ray emission from the Sun in the wavelength range of interest has two components: one of photospheric origin, which dominates in the total SED down to approximately 150 nm, and another which originates in the outer stellar atmosphere (chromosphere and corona), which is composed of optically thin plasma, and which dominates the total SED for wavelengths shorter than 150 nm.

Abundant high-resolution data on the X-ray and UV SED of $\kappa^1$ Cet are available [Güdel et al., 1997; Landi et al., 1997; Guinan and Ribas, 2002; Guinan et al., 2003; Ribas et al., 2005; Telleschi et al., 2005]. These were used to generate its emission measure distribution (EMD), shown in figure 1, by the method of Sanz-Forcada et al. [2002]. The EMD shows how much material is emitted from the outer atmosphere of the star as a function of temperature, and was in this case used to generate a noise-free SED, to be used in further modeling.

An EMD of $\kappa^1$ Cet during a flare was created as well, by scaling a solar flare to $\kappa^1$ Cet: the difference in log(EM) and log(T) between the peak emissions of a quiet and flaring Sun was used to define the peak emission for a flare on $\kappa^1$ Cet, while the shape of the EMD was kept the same as for the flaring Sun (figure 1). The EMDs of the quiet and flaring Sun were taken from Peres et al. [2000] and Orlando et al. [2000], respectively.

The EMDs of $\kappa^1$ Cet were discretized using 42 temperature bins to generate the outer atmosphere SED of $\kappa^1$ Cet in a quiet state and during a flare, using the program XSPEC. A MEKAL model [Liedahl et al., 1995, and references therein] was used to generate the X-ray to EUV part of the SED (0.1-80 nm), and an APEC (Astrophysical

Plasma Emission Code) model [Smith et al., 2001] was used for the UV part (80-320 nm). From the obtained SEDs, fluxes in ergs cm$^{-2}$ s$^{-1}$ were calculated for 74 wavelength intervals. These fluxes are the fluxes as they would be observed at Earth, κ$^1$ Cet being at a distance of ~9.16 pc from Earth. Flux values were rescaled to give the fluxes that would be observed if κ$^1$ Cet were located at the distance of the Sun, and further converted to units of Wm$^{-2}$nm$^{-1}$.

The photospheric emission behaves largely like a blackbody, and at the photospheric temperature of a Sun-like star, the emission for wavelengths shorter than 130 nm is negligible compared to the coronal emission in this range, and was therefore ignored. For longer wavelengths, the photosphere is largely responsible for all the emission in case of the present-day Sun. Since the overall luminosity of the Archean Sun was approximately 75% of the present-day luminosity, its photospheric emission was approximated by taking 75% of the present-day solar radiation in the 130-320 nm range [data from Gueymard, 2004]. The modeled photospheric emission was added to the emission modeled with κ$^1$ Cet, to give the total spectral irradiance of the Sun 4-3.5 Ga as shown in figure 2.

*2.2. Earth's atmosphere 4-3.5 Ga*

Several compositional models were examined to investigate the sensitivity of the spectral irradiance at the surface to variations in concentrations of carbon dioxide, methane, water vapor, and ozone. As $N_2$ is not likely to have evolved much with time, its density profile in the current atmosphere was used, as given by the MSIS-90 model [Hedin, 1991]. The surface concentration of $N_2$ is in this case $2.09 \times 10^{19}$ cm$^{-3}$. For other species, reasonable surface concentrations were taken from the literature (see table 1). Their distribution with altitude was taken proportional to that of $N_2$ for the first 140 km. For higher altitudes, where the slope of the curve for $N_2$ is different, the slope was adjusted for each species, depending on its scale height. The main density profiles that were used are displayed in figure 3. The most frequently used models are listed in table 2.

*2.3. Extinction in the atmosphere*

The extinction in the atmosphere due to absorption and Rayleigh scattering was calculated to first order using the Beer-Lambert law for 0 zenith angle, given by:

$$I(z,\lambda) = I(\infty,\lambda) \cdot \exp[-\sigma_n(\lambda) \cdot \int_z^\infty n_n(s)ds] \qquad (1)$$

where: $I(z, \lambda)$ = intensity of the solar flux at altitude $z$ and wavelength $\lambda$
$I(\infty, \lambda)$ = intensity of the solar flux at the top of the atmosphere and wavelength $\lambda$
$\sigma_n(\lambda)$ = extinction cross-section of species $n$, for a photon at wavelength $\lambda$ = absorption cross-section + Rayleigh scattering cross-section
$n_n(s)$ = density of species $n$ at altitude $s$

The absorption cross-sections that were used are shown in figure 4, and Rayleigh scattering cross-sections were calculated using the standard equation:

$$\sigma_{Rayl.scat.} = \frac{24\pi^3}{\lambda^4 N^2}\left|\frac{m^2-1}{m^2+2}\right|^2 \qquad (2)$$

where: *m* = refractive index of air at standard pressure and temperature
       *N* = number density at standard pressure and temperature

Calculations were carried out iteratively over atmospheric layers of 10 km, starting at an altitude of 400 km.

Simply applying (1), would result in all scattered radiation being lost, while it is actually merely redirected. Therefore we assume that 50% of the scattered radiation proceeds in the forward direction as diffuse radiation. The diffuse radiation generated over each layer in the atmosphere is added to the flux at the top of this layer, and from this the flux at the bottom of the layer is calculated using (1).

This procedure accounts for the major effects of Rayleigh scattering, ignoring smaller effects of longer path length (resulting in a slight overestimate of fluxes), and assuming that all scattered radiation is lost after a second scattering event (resulting in a slight underestimate of fluxes). Effects of the longer path that scattered radiation travels through the atmosphere are expected to be small, as Rayleigh scattering phase functions are biased towards the forward and backward directions, leading to mostly small scattering angles. It is also a fair approximation to assume that scattered radiation is lost after the second scattering event, since only 25% of the originally scattered radiation would proceed in the forward direction, and part of this would be scattered again (or absorbed).

Effects of multiple scattering, re-emission of radiation, scattering by aerosols and clouds, or effects of zenith angles different from 0 degrees were not included in our calculations. We address these issues briefly below, and discuss in section 4.2 in more detail how they affect our results.

Multiple scattering effects become important at optical depths higher than 1, and will result in lower fluxes at the surface than estimated here. However, optical depths greater than 1 are only reached for the case studies with very high concentrations of $CO_2$. In these cases it should be kept in mind that our surface flux estimates may be too high, and therefore represent a worst-case scenario.

Absorption of short-wavelength radiation may result in re-emission of longer wavelength radiation, which could increase radiation levels in the UV. Smith et al. [2004] discuss this issue in detail, and estimate that, depending on the atmospheric composition, up to 10% of the incident short-wavelength radiation (X-rays and γ-rays) can reach the surface as re-emitted UV radiation.

On early Earth, aerosol concentrations may have been relatively high due to more active volcanism, and Brogniez et al. [2002] show that close to the surface (0-30 km), extinction by aerosols can be of comparable importance as Rayleigh scattering at a wavelength of 442 nm. However, the importance of aerosol extinction is much smaller for higher altitudes, and still on early Earth there may have been episodes with low volcanic activity and low aerosol concentrations. Our results are still valid for such episodes, and are probably a slightly less accurate, but still reasonable approximation in situations with higher aerosol concentrations.

Depending on the cloud cover, clouds could shield a significant portion of incoming radiation in the wavelength range considered here. In this work such effects are not included though, and our results therefore apply to cloudless sky situations only.

For our calculations it was assumed that the solar zenith angle was zero degrees. However, as solar zenith angles become larger, radiation will travel a longer path through the atmosphere before reaching the surface, resulting in more scattering and absorption on the way. This causes surface fluxes to be lower for larger zenith angles, up to a factor of ~10 at the poles [see e.g. Cockell, 1998].

## 3. Results

*3.1. Sensitivity analysis: atmospheric models*

In case of the reference $N_2$-$CO_2$ atmosphere, with surface concentrations of $N_2$ and $CO_2$ of $2.09 \times 10^{19}$ and $5.24 \times 10^{18}$ cm$^{-3}$, respectively (80% $N_2$ and 20% $CO_2$ at a total surface pressure of ~1 bar), the short wavelength radiation (~0.1-200 nm) is attenuated already high in the atmosphere, because both $N_2$ and $CO_2$ are strong absorbers in this domain, and scattering cross-sections are high as well. Longer wavelength radiation is progressively less well attenuated, and the longest wavelength radiation (200-320 nm) partially reaches the surface. The surface spectral irradiance in this range is shown in figures 5 and 6, together with irradiances found for other case studies. In figure 6, the surface irradiances from figure 5 are convolved with the DNA action spectrum to show their biological relevance.

The reference $CO_2$ concentration corresponds to a partial pressure of ~0.24 bar, which is in the lower end of the range of $CO_2$ pressures that would be necessary to balance the lower solar luminosity according to Kasting [1993]. Two other tests, with $CO_2$ concentrations 5 times larger and 10 times smaller were performed as well. In the latter case, the presence of another greenhouse gas would be required to balance the lower solar luminosity.

The most important differences between results for different $CO_2$ concentrations occur in the longer wavelength part of the spectrum (> 200 nm), as shorter wavelength radiation is still effectively attenuated by the atmosphere in all cases. For the smallest $CO_2$ concentration, a significant portion of the longer wavelength radiation (> 200 nm) is transmitted, resulting in surface fluxes of ~$10^{-5}$ Wm$^{-2}$nm$^{-1}$ at 200 nm, to ~0.4 Wm$^{-2}$nm$^{-1}$ at 320 nm. For the highest $CO_2$ concentration, much more radiation is attenuated, resulting in surface fluxes of only ~$10^{-16}$ Wm$^{-2}$nm$^{-1}$ at 200 nm, to 0.02 Wm$^{-2}$nm$^{-1}$ at 320 nm.

We showed that the presence of methane or water vapor, in the typical concentrations listed in table 1, has only a minor influence on the surface spectral irradiance. Several case studies with different ozone concentrations were performed as well, as ozone is known to be an important absorber in the present atmosphere. However, ozone concentrations for the Archean, as calculated by Canuto et al. [1982], are much lower than ozone concentrations in the present atmosphere, and even the highest proposed concentration resulted in minor effects on the surface irradiance only.

In none of the above cases, radiation with a wavelength shorter than approximately 200 nm could reach the surface. A few more tests with very tenuous $N_2$-$CO_2$ atmospheres were carried out (10 and 100 times reduced concentrations of both $N_2$ and $CO_2$), but even in these cases only slightly shorter wavelengths could reach the surface due to the high absorption cross-sections of $N_2$ and $CO_2$ in this range, and high scattering cross-sections. In case of a very dense $CO_2$ atmosphere (equivalent to 5 bar), as may have existed on Mars in its early history, much more radiation in the longer wavelength ranges is attenuated, and very little is transmitted.

*3.2. Sensitivity analysis: varying solar fluxes*

Fluxes emitted by the Sun may have been somewhat different from what was modeled, due to a general spread in activity level among (solar-like) stars of the same age [see e.g. Favata and Micela, 2003]. Besides, the Sun may have experienced activity cycles, as it does today. The combined effect of these two factors is estimated to result at most in approximately 5 times higher or lower emitted fluxes in the short wavelength part of the spectrum (< 150 nm) than presented here, causing 5 times higher or lower surface fluxes. Effects for longer wavelengths are estimated to be smaller.

The occurrence of a flare can give rise to very high emitted fluxes over a short period of time. However, the effect of a flare on the total emission from 0.1 to 320 nm is most pronounced in the shorter wavelength part of the spectrum (figure 2), which is very effectively attenuated by Earth's atmosphere. The longer wavelengths (> 200 nm) are not substantially affected by the flare, and it was shown that radiation levels do not change in this part of the spectrum.

*3.3. Comparison with present Earth*

The surface fluxes received at Earth's surface at present were computed following the same procedure as for the Archean. The present solar irradiation was taken from Gueymard [2004], and the atmosphere of the present Earth was approximated by an atmosphere consisting only of $N_2$, $O_2$, and $O_3$. The density profiles of $N_2$ and $O_2$ were based on the MSIS-90 model [Hedin, 1991], and the $O_3$ profile was based on data from McPeters et al. [1999].

Modeled fluxes at the Earth's surface for a wavelength of ~300 nm can be compared to measured values. The modeled values found here (a flux of $1.98 \cdot 10^{-3}$ $Wm^{-2}nm^{-1}$ for the 295-300 nm interval and one of $8.66 \cdot 10^{-3}$ $Wm^{-2}nm^{-1}$ for the 300-305 nm interval) are in good overall agreement with fluxes reported by Nunez et al. [1994], Bernhard et al. [1997], and Piacentini et al. [2002].

To show the importance of the ozone layer in shielding Earth's surface from (part of) the Sun's radiation, one case study was carried out for a present-day atmospheric composition as above, but without ozone. Radiation levels are much higher in this case, and shorter wavelengths reach the surface.

Comparing the surface spectral irradiance at present with results for the early Earth shows that much less radiation, and less energetic radiation, reaches the surface at present than in the Archean for wavelengths of 200 to 320 nm. Also, the spectral response of DNA to the prevailing radiation is much lower at present than during the Archean (figure 6). Figure 7 presents a further quantification of the difference between the two cases, showing the ratio of the Archean weighted surface fluxes over the present-day ones for wavelengths longer than 200 nm. The ratio is extremely high where the absolute values for the weighted fluxes are very low in the present case (< 300 nm). High ratios are easily obtained this way. However, for the longest wavelengths (> 300 nm), where fluxes are substantial at present, Archean weighted surface fluxes can still be more than 12 times higher (300-305 nm; reference atmosphere). For increasing wavelengths the difference decreases, to only 2.4 times higher fluxes for the 315-320 nm interval (again reference atmosphere).

*3.4. Early Mars*

The situation on early Mars may have been very different from the situation at early Earth, for instance because $CO_2$ concentrations may have been much higher. One of our case studies was devoted to early Mars, assuming such a high $CO_2$ concentration (equivalent to ~5 bar). It was found that radiation levels would have been higher than on the present Earth for wavelengths shorter than ~285 nm, but levels would have been much lower for longer wavelengths (see figure 5 and 6).

**4. Discussion and conclusions**

*4.1. Sensitivity analyses*

Short wavelength radiation (< 200 nm) is very effectively attenuated, even by 10 or 100 times lower concentrations of $N_2$ and $CO_2$ than are considered reasonable. Therefore it can be concluded that it is very unlikely that such radiation could reach the surface at any time in Earth's history.

The longer wavelength radiation (> 200 nm) is strongly affected by the amount of $CO_2$ that is present in the atmosphere. In case of the highest $CO_2$ concentration ($2.62 \cdot 10^{19}$ cm$^{-3}$ at the surface) radiation is attenuated relatively strongly, while in case of the lowest $CO_2$ concentration ($5.24 \cdot 10^{17}$ cm$^{-3}$ at the surface) a large portion above 200 nm is transmitted (figure 5). Consequently, the range of surface fluxes that possibly existed in the Archean is still quite broad.

The surface radiation levels presented by Cockell [1998] are higher than found here, even for the case with the lowest $CO_2$ concentration. When scattering is switched off in our model, we do find similar results for this case, so possibly Cockell [1998] did not take this effect into account. Another explanation could be that he used lower $CO_2$ concentrations, but this would not be in agreement with the general view that $CO_2$ concentrations in the Archean were probably high to balance the lower luminosity of the Sun.

Effects of the presence of methane, water vapor, and ozone in concentrations that are likely for the Archean are negligible. Other minor constituents of Earth's early atmosphere, such as ammonia or oxygen, are then also expected to have negligible effects only, as their proposed concentrations are even lower than that of methane (see table 1), and their extinction cross-sections are not high enough to compensate for that.

Surface fluxes found for a present-day atmosphere without ozone are lower than the fluxes found for the Archean reference case. Because there is only a small difference in solar emission between these two cases for the 200-320 nm range, the difference in surface flux must be caused by the difference in atmospheric composition. Where the Archean atmosphere contains, besides ~80% $N_2$, ~20% $CO_2$, the present atmosphere contains ~20% oxygen. As oxygen has a higher absorption cross-section than $CO_2$ for the relevant wavelength range (200-320 nm), it follows that the present atmosphere has more absorbing power, even without ozone, than the Archean atmosphere.

The occurrence of a modeled flare left the surface spectral irradiance unaffected, and wavelength independent variations in activity level, both due to an intrinsic spread in activity between stars of the same age and due to possible activity cycles, are expected to cause at most 5 times larger or lower surface fluxes for the shortest wavelengths (< 200 nm), that do not reach the surface. The longer wavelengths that are emitted by the photosphere are much less affected.

Still, even a change of a factor 5 in surface flux can be considered small compared to the effects of different $CO_2$ concentrations in the atmosphere. The surface irradiance from 200 to 320 nm is hence much more strongly influenced by plausible differences in atmospheric composition ($CO_2$ concentration) than by differences in solar emission.

*4.2 Effects of assumptions and approximations*

Re-emission of X-rays and γ-rays would cause UV fluxes to increase. Using the results of Smith et al. [2004], we can crudely estimate that UV fluxes arising from redistributed X-rays could be of the order of ~$10^{-6}$-$10^{-5}$ $Wm^{-2}nm^{-1}$, assuming that 1-10% of the average X-ray flux incident at the top of the atmosphere (~$10^{-4}$ $Wm^{-2}nm^{-1}$, see figure 2) is redistributed in this wavelength domain. Redistribution of γ-rays would still increase this estimate.

Cases with high extinction (high $CO_2$ concentration), and therefore relatively low radiation levels, could significantly be affected by re-emission, depending on the wavelengths at which radiation is most effectively re-emitted. Cases with low extinction, for which surface radiation levels are already much higher, would probably not be affected much.

Extinction by aerosols and effects of zenith angles different from 0 degrees would tend to decrease surface radiation levels. Extinction by aerosols could be comparable to the effect of Rayleigh scattering [Brogniez et al., 2002], which reduces surface levels by a factor of ~10 at a central wavelength of 250 nm (this followed from test cases for which scattering was switched off). The combined effect of aerosols and other zenith angles can therefore possibly reduce surface levels up to a factor of 100 at 250 nm at the poles, and even more for shorter wavelengths (and somewhat less for longer wavelengths). Clouds, if present, could provide even additional shielding. For these reasons, our results should be interpreted to represent the worst-case environment, in particular for the cases with low extinction, where re-emission of X-rays and γ-rays is expected to counterbalance the above effects only slightly.

*4.3. Early Mars*

We have shown that surface radiation levels on early Mars would have been very low in case of a dense (~5 bar) $CO_2$ atmosphere. Still, it is uncertain whether Mars indeed had such a dense atmosphere, and if it did not, radiation levels may have been much higher, unless shielding was provided by something else.

A possible other shielding substance would be $SO_2$, since $SO_2$ has high absorption cross-sections in the longer wavelength ranges (150 to 320 nm), comparable to $O_3$ (figure 4). There are no indications that $SO_2$ had high concentrations in early Earth's atmosphere, but on Mars it may have been much more abundant, as suggested by recently discovered sulphate deposits [Gendrin et al., 2005]. However, if liquid water was present at Mars at the same time, it is unlikely that high concentrations of $SO_2$ could build up in the atmosphere due to its high solubility [Kasting et al., 1989]. It is therefore uncertain whether Mars' surface could have been shielded from harmful radiation by an atmosphere enriched in $SO_2$.

*4.4. Implications for life on Earth*

It can be concluded from our results that any life that would have evolved at Earth's surface 4-3.5 Ga, would have been exposed to damaging radiation fluxes (figure 6). It would not only have experienced higher levels of solar radiation in general, but also more damaging (shorter wavelength) radiation than now, and therefore it is questionable if life would have been sustainable at all at Earth's surface in the Archean. It may be more realistic to envisage an origin of life in an environment where it was shielded from the harmful radiation, for instance in the deep sea. This is sometimes called the Berkner-Marshall hypothesis [Berkner and Marshall, 1965].

Cockell [2000b] modeled how much radiation could penetrate the Archean oceans as a function of water depth. Assuming an atmosphere similar to the one in this study with the smallest $CO_2$ concentration, he found that DNA damage rates were perhaps three orders of magnitude higher in the surface layer of the Archean ocean than in the present-day oceans. Only at a depth of ~30 m damage may have been similar to the surface of the present-day oceans.

However, even though life may have formed in the sheltering environment of the sea, at some point, life did emerge at Earth's surface as it started to use sunlight as a source of energy. From the results of this study, it can not be expected that the conditions at Earth's surface became less hostile shortly after 3.5 Ga. Only major changes in either solar irradiation or atmospheric composition could have caused a significant change in the surface spectrum. Solar irradiation only decreased with time for the short wavelength range (0.1-150 nm) that is effectively filtered by all atmospheric compositions. Solar emission increased slightly with time for the longer wavelengths (200 to 300 nm) that can reach the surface, rather resulting in a (minor) harshening of the environment than in a relaxation of the conditions. Atmospheric composition, which is of major importance for the surface conditions, changed dramatically only around 2.2-2 Ga when concentrations of oxygen, and thereby ozone, started to build up [see e.g. Rye and Holland, 1998]. The evolution of $CO_2$, another important component, is uncertain: it is not clear how high its concentration was in the Archean, though probably higher than today, and how and when it evolved to its present state. In any case, its concentration is rather believed to have decreased with time than to have increased, which would then again have resulted only in still tougher conditions. For these reasons, any life that appeared at Earth's surface before 2.2-2 Ga, was probably still subject to harmful solar radiation levels.

**Acknowledgements**

We appreciate the comments from J.F Kasting, A. Segura and E.F. Guinan on the original manuscript, and we thank ESA, and in particular RSSD, for the support provided.

**Figure captions and tables**

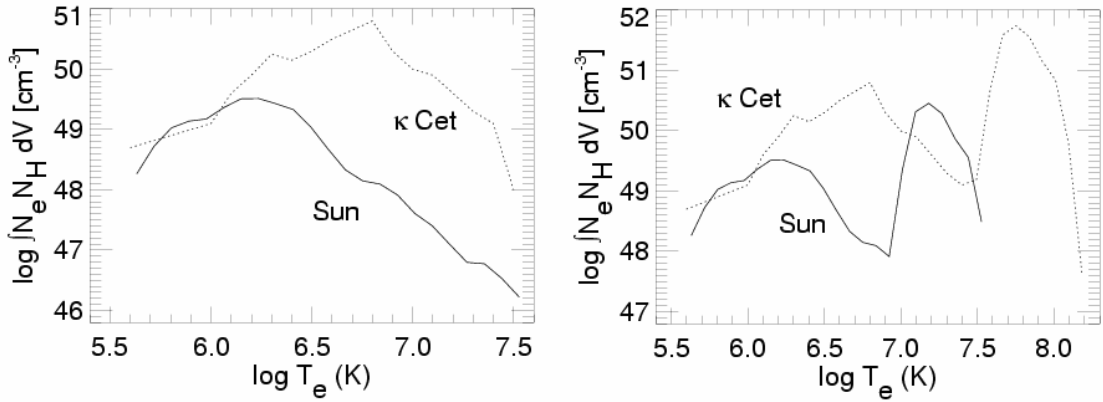

Figure 1. The emission measure distributions (EMDs) of the Sun and κ[1] Cet in quiet state (left) and during a flare (right). κ[1] Cet emits more radiation, and at higher temperatures than the Sun, resulting in higher fluxes in particular in the shorter wavelength (higher energy) domain.

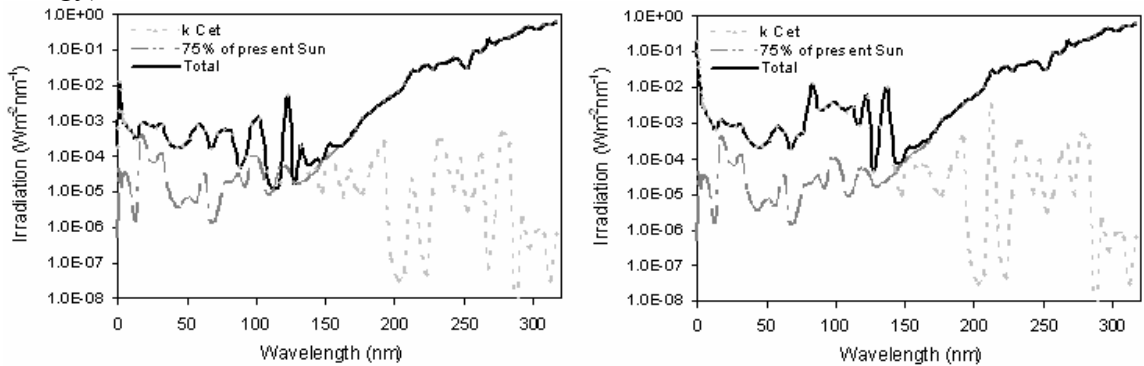

Figure 2. The modeled spectral irradiance of the Sun 4-3.5 Ga in quiet state (left) and during a flare (right). The total irradiance consists of the emission from the corona and the transition region, represented by the SED modeled for κ[1] Cet, and the emission from the photosphere, represented by 75% of the total present-day solar irradiation. Shorter wavelength (< 150 nm, higher energy) radiation is mainly emitted by the corona and the transition region, while the photosphere is largely responsible for the longer wavelength emission.

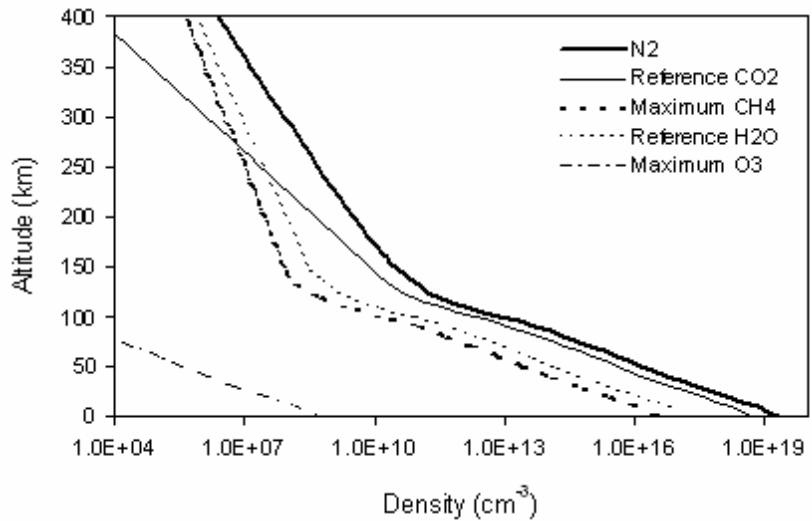

Figure 3. Main density profiles used for atmospheric models.

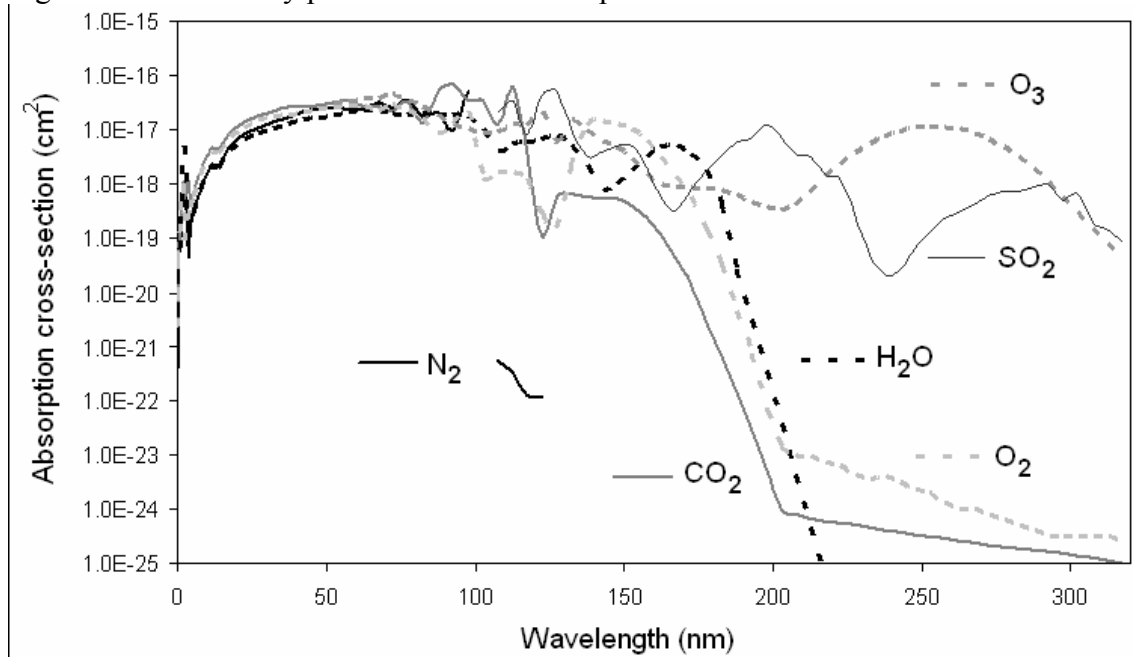

Figure 4. Absorption cross-sections for $N_2$, $CO_2$, $CH_4$, $H_2O$, $O_3$, $O_2$, and $SO_2$. Data are from Ditchburn [1955], Sun and Weissler [1955], Ogawa and Cook [1958], Thompson et al. [1963], Schürgers and Welge [1968], Huffman [1969], Ackerman [1971], Shemansky [1972], Mount et al. [1977], Manatt and Lane [1993], Yoshino et al. [1996], Avakyan et al. [1998], Bogumil et al. [2003] and Fillion et al. [2004]. In some cases, where data were not available, extrapolations were used, or cross-sections were assumed zero, where values could be expected to be very small.

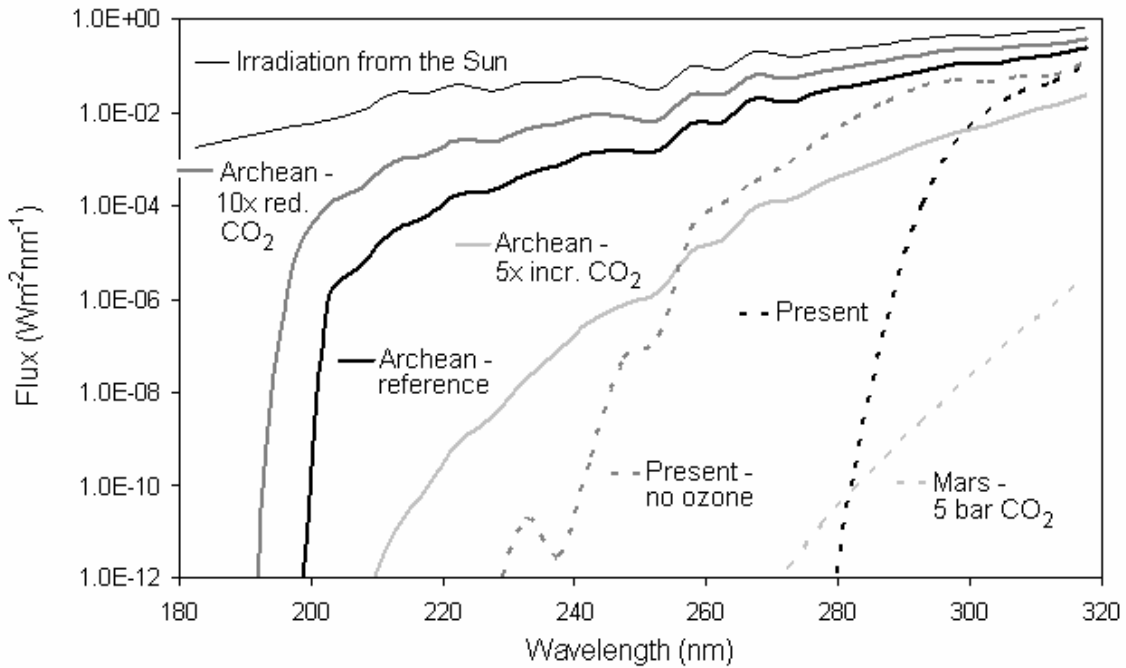

Figure 5. Irradiation at the top of the atmosphere of the Archean Earth, and the surface irradiance for several case studies: the Archean reference atmosphere and varying $CO_2$ concentrations, the present Earth with and without ozone, and Mars in case of a very dense $CO_2$ atmosphere (equivalent to 5 bar). Only the interval from 180 to 320 nm is displayed, as shorter wavelengths do not reach the surface.

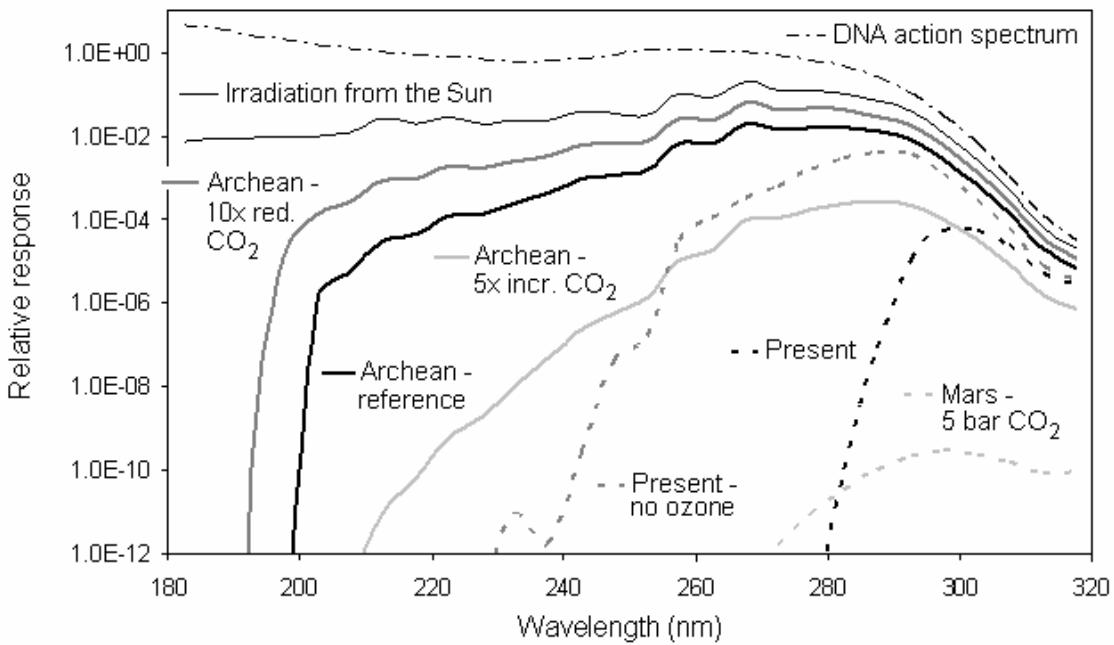

Figure 6. Irradiance at the top of the atmosphere and at the surface, weighted by the DNA action spectrum. Action spectra provide a relative measure of a certain biological response as a function of wavelength, and weighting the surface irradiances shown in figure 5 by the action spectrum of DNA therefore gives an indication of their biological

significance. The action spectrum used here is normalized to a value of 1 at 265 nm, and is based on data from Setlow (1974) and Lindberg and Horneck (1991).

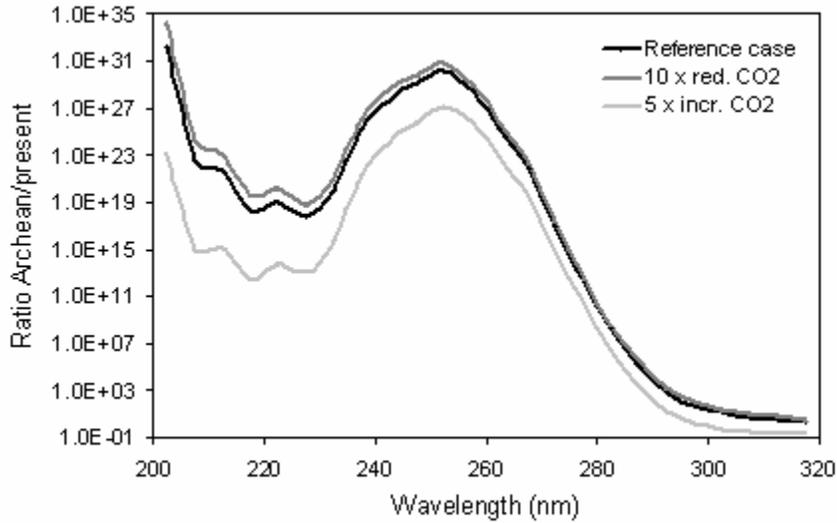

Figure 7. Ratio of the weighted surface irradiance in the Archean over the present-day irradiance for three different atmospheres: the reference atmosphere, an atmosphere with a 10 times smaller $CO_2$ concentration, and an atmosphere with a 5 times larger $CO_2$ concentration. The ratios between Archean and present irradiances are very high. However, absolute values for the present irradiance are very low, especially for wavelengths shorter than 300 nm, so that high ratios are easily obtained.

| Component | Surface concentration in $cm^{-3}$ | Relevant references |
|---|---|---|
| $N_2$ | $2.09 \times 10^{19}$ | Selsis (2004) |
| $CO_2$ | $5.24 \times 10^{18}$, $5.24 \times 10^{17}$, $2.62 \times 10^{19}$ | Kasting (1993), Canuto et al. (1982) |
| $H_2O$ | $2.2 \times 10^{17}$ | Canuto et al. (1983) |
| $CH_4$ | $6.6 \times 10^{15}$, $4.0 \times 10^{16}$ | Catling et al. (2001) |
| $NH_3$ | $4.2 \times 10^{13}$ | Bada and Miller (1968) |
| $O_2$ | $2.2 \times 10^{8}$ | Kasting (1993), Canuto et al. (1982) |
| $O_3$ | $3.5 \times 10^{3}$, $3 \times 10^{6}$, $6 \times 10^{8}$ | Canuto et al. (1982) |

Table 1. Proposed concentrations of atmospheric constituents for usage in atmospheric models. Estimates are based on required greenhouse gas concentrations to balance the lower solar luminosity ($CO_2$, $CH_4$), photochemical models ($O_2$, $O_3$), equilibria calculations ($NH_3$), or on similarity with the present atmosphere ($N_2$, $H_2O$).

| Model | Surface concentration ($cm^{-3}$) | | | | $P_{surface}$ (bar) |
| | $N_2$ | $CO_2$ | $O_2$ | $O_3$ | |
|---|---|---|---|---|---|
| Archean - reference | $2.09 \times 10^{19}$ | $5.24 \times 10^{18}$ | - | - | 1 |
| Archean - 10 x red. $CO_2$ | $2.09 \times 10^{19}$ | $5.24 \times 10^{17}$ | - | - | 0.8 |
| Archean - 5 x incr. | $2.09 \times 10^{19}$ | $2.62 \times 10^{19}$ | - | - | 1.8 |
| CO2 | | | | | |
| Mars - 5 bar CO2 | $2.09 \times 10^{19}$ | $1.05 \times 10^{20}$ | - | - | 5 |
| Present | $2.09 \times 10^{19}$ | - | $5.62 \times 10^{18}$ | $1.35 \times 10^{12}$ | 1 |
| Present – no ozone | $2.09 \times 10^{19}$ | - | $5.62 \times 10^{18}$ | - | 1 |

Table 2. Main atmospheric models that were used in the analysis of the sensitivity of the surface irradiance to atmospheric composition.